\documentstyle[12pt]{article}
\def\lqq{\lq \lq }

\begin{document}

\begin{titlepage}

\rightline{Preprint  IFUNAM FT 94-43}
\rightline{March 1994}
\rightline{May 1994 (corrected)}
\begin{center}
{\large Two electrons in an external oscillator potential: hidden algebraic
structure
\vskip 0.5cm
by
\vskip 0.5cm
 Alexander Turbiner}$^{\dagger }$
\vskip 0.5cm
Instituto de Fisica, UNAM, Apartado Postal 20-364, 01000 Mexico D.F., Mexico
\end{center}

\begin{center}
{\large ABSTRACT}
\end{center}
\vskip 0.5 cm
\begin{quote}
It is shown that the Coulomb correlation problem for a system of two electrons
(two
charged particles) in an external oscillator potential possesses a hidden
$sl_2$-algebraic structure being one of  recently-discovered
quasi-exactly-solvable
problems. The origin of existing exact solutions to this problem, recently
discovered
by several authors, is explained.  A degeneracy of energies in
electron-electron
and electron-positron correlation problems is found. It manifests the first
appearence
of hidden $sl_2$-algebraic structure in atomic physics.
\end{quote}

\vfill

\begin{center}
PASC: 03.65.Fd and 03.65.Ge
\end{center}

\noindent
$^\dagger$On leave of absence from the Institute for Theoretical
and Experimental Physics,
Moscow 117259, Russia\\E-mail: turbiner@teorica0.ifisicacu.unam.mx
or turbiner@vxcern.cern.ch

\end{titlepage}

\newpage
The problem of evaluation of effects of inter-electronic interactions is one of
the central
problems in atomic physics. The main difficulty comes out from the fact that
this problem
cannot be solved exactly even in particular cases, while numerical solutions
are too
complicated to gain a proper intuition. Therefore, it is quite important to
find and elaborate situations, where this problem can be modelled in some
relevant way, admitting exact,
analytic solutions. One of such situations has been described recently in
\cite{sg,taut}.
A system of two electrons in an external harmonic-oscillator potential
with an additional linear interaction was studied in the relative coordinate,
defined by the Hamiltonian
\footnote{ In \cite{mosh} so called pseudo atoms were introduced : quantum
system
made from atomic ones in which all Coulomb interactions are replaced by
oscillator ones,
attractive or repulsive as the case may be. The system described by (1) can be
treated as modified two-electron pseudo atom: where Coulomb attractive
interactions are replaced by
oscillator ones, while Coulomb repulsion remains unchanged or modified by the
linear interaction.}
\begin{equation}
H = -\nabla^2_1 + \omega^2r^2_1 -\nabla^2_2 + \omega^2r^2_2 +
{2\beta \over |{\bf r}_1-{\bf r}_2|} + \lambda |{\bf r}_1-{\bf r}_2|
\end{equation}
where $r_{1,2}$ are the coordinates of the electrons and $\beta=1$. Atomic
units $\hbar=m=e=1$ are used throughout and an overall factor ${1 \over 2}$ is
omitted.
It was found that for certain values of oscillator frequency $\omega$ and the
parameter
$\lambda$ some eigenstate of (1) can be obtained analytically.
The main purpose of this note is to show that this feature is nothing but a
consequence of the fact that (1) is one of recently-discovered
quasi-exactly-solvable Schroedinger operators \cite{t1,t2}. It implies an
existence of a hidden algebraic structure \cite{t3}.
Hereafter, we will focus on the case $\lambda=0$.

Quasi-exactly-solvable problems are quantum-mechanical problems for which
several eigenstates can be found explicitly. They occupy an intermediate place
between exactly-solvable (like Coulomb potential, harmonic oscillator etc) and
non-solvable.
The quasi-exactly-solvable Schroedinger equations appear in two forms:
(i) the Hamiltonian with an infinite discrete spectrum with several eigenstates
known algebraically and (ii) the Hamiltonian depending on a free parameter, say
$\beta$,
and a certain fixed magnitude of energy corresponds to the $i$th state of the
Hamiltonian
at $i$th value of parameter $\beta$ (where $i=0,1,2,\ldots, n$)
\footnote{Precisely speaking, this means that for the parameters
$\beta_0, \beta_1, \ldots, \beta_n$ , the ground-state energy at $\beta_0$ is
equal to
the energy of the first-excited state at $\beta_1$,is equal to the energy of
2th-excited state
at $\beta_2$ etc, is  equal to the energy of $n$th excited state at
$\beta_n$.}.
Those problems are named the first- and the second type, respectively.
Surprisingly, exactly-solvable problems like the Coulomb problem, the Morse
oscillator,
the P\"{o}schl-Teller potential have two equivalent representations,
either as the first-type problems or as the second-type ones \cite{t1}
\footnote{For the Coulomb the second-type representation is nothing but the
well-known Sturm representation.}.

The underlying idea behind quasi-exactly-solvability is the existence of a
hidden
algebraic structure. Let us recall a general construction considering the
one-dimensional Schroedinger equation as an example. Take the algebra $sl_2$
realized in the first-order differential operators
\[ J^+_n = r^2 d_r - n r,\  \]
\begin{equation}
 J^0_n = r d_r - {n \over 2} \  ,
\end{equation}
\[ J^-_n = d_r \ ,  \]
\noindent
where $r \in R$ and $d_r \equiv {d \over dr} $. Those three generators obey
$sl_2$ --algebra commutation relations for any value of the parameter $n$. If
$n$ is a non-negative
integer, the algebra (2) possesses $(n+1)$-dimensional irreducible
representation
\begin{equation}
{\cal P}_{n+1}(r) \ = \ \langle 1, r, r^2, \dots , r^n \rangle \ .
\end{equation}
It is evident,  that taking any polynomial in the generators (2), we arrive at
a differential
operator having the space (3) as the finite-dimensional invariant subspace.
In other words, almost any polynomial in the generators (2) possesses $(n+1)$
eigenfunctions in the form of polynomial in $r$ of degree $n$.

Let us take the quasi-exactly-solvable operator
\footnote{It belongs to the case VIII in the classification \cite{t2}.}
\begin{equation}
 T_2 =  -J^0_n J^-_n + 2\omega_r J^+_n  - (n/2+2l+2) J^-_n - \omega_r n \ .
\end{equation}
Substituting (2) into (4), one gets the differential operator
\begin{equation}
T_2 (r,d_r;n)= -rd_r^2 + 2(\omega_r r^2 - l -1)d_r - 2\omega_r n r
\end{equation}
for which one can define the spectral problem
\begin{equation}
T_2 (r,d_r;n) p(r) = -\beta (n) p(r)
\end{equation}
where $\beta (n)$ is a spectral parameter. It is clear, that this problem
possesses
$(n+1)$ eigenfunctions, $p_0 (r),p_1(r),p_2(r),\ldots, p_n(r)$ in the form of
polynomial
of the $n$-th power. Other eigenfunctions are non-polynomial and
in general, they cannot be found in closed analytic form.
Now let us make a gauge transformation in (5),(6), introducing a new function,
\begin{equation}
u(r)= r^{l+1} p(r) \exp (-\sqrt{\omega_r}r^2/2) ,
\end{equation}
then make a replacement in the last term in (5):
\begin{equation}
2\omega_r n = \epsilon' - \omega_r (2l+3)
\end{equation}
where $\epsilon'$ is a new parameter,  and divide (6) over $r$.
Finally, we obtain the equation
\begin{equation}
[ -d^2_r + \omega^2_r r^2 + { \beta \over r} + {l(l+1) \over r^2}] u(r) =
\epsilon' u(r)
\end{equation}
\noindent
Putting in (9) $\beta =1$ and saying that now a new spectral parameter is
$\epsilon'$,
we arrive at the equation (9) of the paper \cite{taut}. If $\omega_r=2\omega$
and $2\epsilon'=\epsilon$ is the energy of the relative motion, this equation
appears
in \cite{taut} as a radial equation for the relative motion  in (1) after
separation
of the c.m. motion.

Equation (9) is a particular case of the quasi-exactly-solvable Schroedinger
equation of the second type (Case VIII in the classification  \cite{t2})
\footnote{ General case VIII corresponds to $\lambda \neq 0$. In \cite{t1} this
problem was named the {\it generalized Coulomb problem.}}.
{}From the physical viewpoint, the parameter $\beta$ in (1) has a meaning of
the
constant of the inter-electronic  interaction. This parameter can be changed by
replacing an electron by a charged particle with charge $Z$. In principle,
keeping the frequency $\omega$ and the energy $\epsilon'$ fixed, for any $n$
and $l$ one can find $(n+1)$ systems of two particles with different charges in
the
oscillator potential related each other via hidden $sl_2$ algebraic structure
(see a discussion in footnote 2).

Now let us describe some features of the eigenvalue problem (6).
\begin{itemize}
\item{(i)}. It is clear that the operator
$T_2$ is self-adjoint and hence its eigenvalues are real. The first $(n+1)$
eigenvalues
coincide with the eigenvalues of the Jacobian matrix with vanishing diagonal
matrix
elements
\newpage
\[
\hat H =
\]
\begin{equation}
\!\!\!\!\!
\left ( \begin{array}{ccccccc}
0  & 2 \omega_r & 0 & \cdots & 0 & 0 & \cdots  \\
n(n+1+2l)   & 0 & 4\omega_r & \cdots & 0 & 0 &  \cdots  \\
\vdots & \vdots &   & \vdots &  & \vdots &\vdots \\
0 & \cdots & (i+1)(i+2+2l) & 0 & 2(n-i+1) \omega_r &\cdots & 0 \\
\vdots  & \vdots   &  & \vdots &  & \vdots  & \vdots   \\
0 &  0 &  \cdots & 0 & \cdots &  0 & 2n \omega_r  \\
0 & 0 &  \cdots & 0 & \cdots & 2(l+1) & 0
\end{array} \right )
\end{equation}
One can show that the spectrum of (10) is symmetric :
\begin{equation}
S_{n+1} (\beta) = (-1)^n S_{n+1} (-\beta)\ ,\quad S_{n+1} (\beta) \equiv
\det{||\hat H - \beta||}
\end{equation}
that follows from the fact that all odd powers of the matrix $\hat H$ are
traceless,
$tr \hat H^{2j+1}=0, j=0,1,\ldots$
\footnote{I am grateful to P. Mello for a discussion of this point.}. So, for
any $\omega_r>0$ there exist \linebreak $[(n+1)/2]$
\footnote{$[a]$ means integer part of $a$.}
 positive eigenvalues
and the same amount negative eigenvalues.
This property leads to an important conclusion: for the fixed $n, l$ and
$\omega_r$ there
exist two eigenstates, one at $\beta >0 $ and another at $\beta <0$, degenerate
in energy
(see (8)). In particular, this may allow  the electron-electron correlation
energy to be to
related with the electron-positron one in the problem (1) (see discussion
below).

\item{(ii)}. Another important property of (6) is  that all eigenvalues
$\beta^{(n)} \propto \sqrt {\omega_r}$ and, hence, depend on $\omega_r$
monotonously.
For instance,
\[
\beta^{(1)}_{\pm} = \pm 2 \sqrt {\omega_r (l+1)}
\]
\[
\beta^{(2)}_{\pm, 0} = \{ \pm 2 \sqrt {\omega_r (4l+5)} \ ,\ 0 \}
\]
In order to find the eigenvalues of (6) it is enough to perform calculations in
one point on $\omega_r$-axe, e.g. at $\omega_r=1$.
\end{itemize}
The situation becomes slightly more complicated, if we want to keep
the parameter $\beta$ fixed in the formula (1), declaring that now we want to
consider
namely two-electron (or electron-positron) system, which implies $\beta=1
(-1)$.
The relevant formulation of the problem is the following.

Let us fix $n$ and $l$. This defines unambiguously the functional form of the
pre-exponential factor in (7). Take a positive eigenvalue $\beta$ in (6), which
corresponds
to repulsion of the particles in (1)). It depends on the parameter $\omega_r$
monotonously, growing from zero up to infinity, which means that one can always
find the value of
$\omega_r$ for what $\beta =1$ or any positive number.  Since there exist
$[(n+1)/2]$
positive eigenvalues of $\beta$ (see above), each of them is equal to one for a
certain
value of  $\omega_r$. Correspondingly, the lowest eigenvalue (ground state, no
nodes
in $p(r)$ following the oscillation theorem) leads to the smallest value of
$\omega_r$,
the next eigenvalue leads to bigger value of $\omega_r$ (one positive root in
$p(r)$) etc . Finally, we arrive at  $[(n+1)/2]$ values of the parameter
$\omega_r$, for each of them
the problem (1) has the analytic solution of the form (7) with $p(r)$ as a
polynomial of the
$n$th degree with number of positive roots varying from 0 up to $[(n+1)/2]$
(see
Fig.1, where the case $n=3$ is described as an illustrative example)
\footnote{it explains a systematics found in numerical calculation in
\cite{taut}.}.

Taking the negative eigenvalues of $\beta$ in (1) (that corresponds to
attraction of the
particles), one can repeat above considerations with the only difference that
the number of
positive roots varies from $[(n+1)/2]$ up to $n$. Following the property (i)
for any eigenstate
from algebraized part of the spectra (see above) of the problem (1) with
positive $\beta$
one can find an eigenstate with negative $\beta$ with the same energy. For
example,
for the fixed $n$ and the minimal $\omega_r$: $\omega_r^{(0)}$, the ground
state energy
at $\beta =1$ ($p(r)$ has no positive roots) is equal to the the energy of the
$n$th excited
state at  $\beta =-1$ ($p(r)$ has $n$ positive roots). For $\omega_r^{(1)}$,
the energy of
the 1st excited state at $\beta =1$ ($p(r)$ has one positive root) is equal to
the energy of the $(n-1)$th excited state at $\beta =-1$ ($p(r)$has $(n-1)$
positive roots) etc (see e.g. Fig.1).
This is reminiscent of the situation in one-dimensional supersymmetic quantum
mechanics
by Nicolai-Witten, where if the supersymmetry is unbroken, all states of
bosonic sector
(except lowest one) are degenerate with the states of fermionic one.

It is worth noting that recently it was shown \cite{t3} that whenever some
analytic
solutions for eigenfunctions of a certain one-dimensional
(or reduced to one-dimensional) Schroedinger equation occur, it signals
the existence of the hidden algebra $sl_2$. Our present results manifest the
first
appearance of quasi-exactly-solvability in atomic physics. Novel developments
of hidden algebra method in quantum mechanics, solid-state physics and quantum
field theory, and also mathematical foundations can be found in Refs.
\cite{ams}.

I would like to thank C.~Bunge for bringing my attention to the papers
\cite{sg,taut} and useful comments. I am grateful to the Instituto de Fisica,
UNAM for kind hospitality extended to me.

\end{document}